\documentclass[12pt]{article}

\newcommand{\x}{arXiv:}
\newcommand{\m}{\mathrm}
\newcommand{\be}{\begin{equation}}
\newcommand{\ee}{\end{equation}}
\newcommand{\ba}{\begin{eqnarray}}
\newcommand{\ea}{\end{eqnarray}}

\usepackage{graphicx}
\usepackage{amssymb}
\usepackage{amsmath}
\usepackage[T1]{fontenc} %for \boldsymbol
\usepackage[ansinew]{inputenc} %for \boldsymbol
\usepackage[nosort]{cite}
\newcommand{\inbar}{\vrule height1.57ex width.4pt depth0pt}
\newcommand{\SW}{\relax{\hbox{$\ \inbar\kern-.285em{\rm S}$}}}

\begin{document}
\thispagestyle{empty}
\begin{center}

\null \vskip-1truecm \vskip2truecm

{\Large{\bf \textsf{Complexity vs. Vorticity}}}

{\large{\bf \textsf{}}}

{\large{\bf \textsf{}}}

\vskip1truecm

{\large \textsf{Brett McInnes}}

\vskip1truecm

\textsf{\\  National
  University of Singapore}

\textsf{email: matmcinn@nus.edu.sg}\\

\end{center}
\vskip1truecm \centerline{\textsf{ABSTRACT}} \baselineskip=15pt
\medskip
In the study of ``holographic complexity'', upper bounds on the rate of growth of the (specific) complexity of field theories with holographic duals have attracted much attention. Underlying these upper bounds there are inequalities relating the parameters of the dual black hole. We derive such an inequality in the case of the five-dimensional AdS-Kerr black hole, dual to a four-dimensional field theory with a non-zero angular momentum density. We propose to test these underlying inequalities ``experimentally'', by using the conjectured analogy of the field theory with phenomenological models of the Quark-Gluon Plasma. The test consists of comparing data for the parameters of the QGP with the upper bound on the relevant combination of black hole parameters. The bound in the non-rotating case passes the test: in this sense, it is confirmed ``experimentally''. In the rotating case, the inequality makes predictions regarding the entropy density of the vortical plasma, recently observed by the STAR collaboration.

\newpage

\addtocounter{section}{1}
\section* {\large{\textsf{1. Rotation and Complexity in the AdS$_5$ Background}}}
The gauge-gravity duality \cite{kn:nat} postulates an exact equivalence between a gravitational theory in an asymptotically AdS bulk spacetime and a field theory on the conformal boundary of that spacetime. In the case in which the bulk harbours a black hole\footnote{Throughout this work, we consider only four-dimensional boundary theories, and so \emph{all black holes in this work are five-dimensional}.}, one has a detailed understanding of how this duality works for the exterior of the black hole. The interior, however, presents a greater challenge, even regarding the most basic aspects of the geometry. In particular, the simple fact that there is no timelike Killing vector field in that region of spacetime means that the interior structure is strongly dynamic, on time scales very different to those associated with the equilibration of the exterior and the dual system at infinity.

Motivated by this challenge, Susskind and others \cite{kn:suss1,kn:suss2,kn:suss3,kn:suss4,kn:suss5} have proposed a relation between the dynamics of the black hole interior geometry and the growth of the \emph{complexity} in a dual field theory, the point being that the latter can \emph{grow very slowly}. It can in fact grow for a very long time after thermal equilibrium has been established: its growth is therefore a plausible candidate to serve as the dual of some measure\footnote{There are differing views as to what this measure should be: see for example \cite{kn:jacob}. We will not need to take a stand on this question.} of the evolution of the black hole interior.

It is clear that, in order to pursue this idea, one needs a good understanding of the factors controlling the rate of growth of complexity: see for example \cite{kn:carmimyers} for detailed studies of this. In particular, for a close relation between complexity growth and the black hole interior dynamics to work, it is essential that the boundary complexity \emph{should not grow too quickly}. On the other hand, there is evidence \cite{kn:second} to suggest the existence of a ``second law of complexity'', and, from that point of view, one does not want the complexity to grow too slowly. Of course, it will ultimately be important to specify the meanings of ``too quickly'' and ``too slowly'' much more precisely. Our main (though not exclusive) focus here will be on the former.

We therefore investigate the possible existence of an \emph{upper bound} on the rate of growth of the complexity per unit energy or ``specific complexity''\footnote{We focus on the complexity  \emph{per unit energy (or mass)}, for the same reason that, in thermodynamics, one uses the entropy per unit mass (``specific entropy''): this is an intensive quantity. This would be necessary, for example, if one were to discuss field theories dual to AdS black holes with planar horizons (which we will not do in this work, however). We do not consider systems with time-dependent masses or energies. Note that the rate of change of specific complexity has the virtue of being dimensionless in natural units.} of the boundary system. It has been proposed (see \cite{kn:suss3}) that this upper bound can be constructed explicitly in two steps, as follows.

First, there is a heuristic argument suggesting that the rate of growth of the specific complexity, $\hat{\textsf{C}}$, is bounded by a dimensionless multiple of some combination of the thermodynamic parameters of the system. In the simplest cases, this is the \emph{product of the specific entropy of the system with its temperature}.

Second, the natural way to investigate this quantity is to use holography a \emph{second} time (now applying it to the exterior instead of the interior of the black hole), by examining the analogous quantity for an asymptotically AdS thermal black hole dual to the given boundary system. Again, in the simplest cases this is $ST/\mathcal{M}$, where now $S$ is the entropy of the black hole, $\mathcal{M}$ is its physical mass\footnote{This $\mathcal{M}$ is proportional but not equal to the parameter $M$ occurring in black hole metrics. To make the distinction clear, we use natural units based on the femtometre: 1 fm $\approx \left(197.3\, \m{MeV}\right)^{-1}$. In these units, $M$ (in five dimensions) has units of (fm)$^2$, whereas $\mathcal{M}$ (in all dimensions) must have the correct units for mass or energy, (fm)$^{-1}$. Planck units are never used in this work.}, and $T$ is its Hawking temperature. It turns out that, for black holes of a given general type, \emph{there is also an upper bound on this quantity}\footnote{For asymptotically flat Schwarzschild spacetimes, this quantity is a constant (independent of the black hole parameters, in this case, the mass): in natural units, it is 1/2 in four spacetime dimensions, 2/3 in five. In the asymptotically AdS case, however, it is never constant in this sense.}.

Combining these observations, one obtains upper bounds on the rate of growth of the specific complexity of the boundary theory. For example, for boundary field theories corresponding to an asymptotically AdS$_5$ black hole (the examples we consider are AdS$_5$-Schwarzschild, AdS$_5$-Reissner-Nordstr\"om, and AdS$_5$-Kerr black holes), the simplest bound is
\begin{equation}\label{ALPHA}
{d\hat{\textsf{C}}\over dt}\;<\;{2\over \pi}.
\end{equation}
Here $\hat{\textsf{C}}$ is the specific complexity, and $t$ is time according to a stationary asymptotic observer; the numerical factor has been chosen so as to match the factor $2/\pi$ proposed in \cite{kn:suss3}, which itself is motivated by the considerations of \cite{kn:lloyd}.

Note that we are only concerned here with (upper) \emph{bounds} of this kind: the actual \emph{value} of the rate, and its possible dependence on time (particularly at early times) are very much more complex matters; see for example \cite{kn:myersgoto}, and its references, for some of the techniques relevant to those questions. By focusing only on upper bounds, we can avoid these complications.

We have called (\ref{ALPHA}) the ``simplest'' bound because, clearly, there cannot be a \emph{unique} upper bound. One can have various upper bounds which may be more or less restrictive, but also more or less useful; for example, one bound might involve parameters that are easier to determine than those of the other. Two different upper bounds can of course both be valid. For example, (\ref{ALPHA}) can be improved for AdS-Reissner-Nordstr\"om black holes (see \cite{kn:suss3,kn:zhong}) but clearly this does not mean that (\ref{ALPHA}) is not true of them.

Evidently there are two ways of improving (\ref{ALPHA}), corresponding to the two steps in its proof. First, one might try to construct a more sophisticated bound on the rate of growth of complexity than the one given simply by the product of the specific entropy with the temperature: this is the approach of \cite{kn:suss3,kn:zhong}. The second approach is to try to improve the bound on the black hole quantity $ST/\mathcal{M}$. This second approach will be taken in this work. In particular, we will attempt this in the case of \emph{boundary matter with a non-zero angular momentum density}.

Indeed, matter with zero angular momentum is essentially non-generic; for example, the plasmas produced in heavy-ion collisions typically contain vortices, as we will discuss in detail below\footnote{Because we wish to consider the effects of rotation, we confine our attention here to black holes with topologically spherical event horizons. In the holographic context, this means that the boundary field theory is defined on a spacetime with compact spatial sections, which is acceptable provided that the total spatial volume is very much larger than that of the system being described; in other words, if the curvature is negligible. (See \cite{kn:nat}, Chapter 14, for a discussion.) Of course we must verify this; see Section 5 below. Earlier holographic applications of (four-dimensional) AdS-Kerr black holes with compact horizons include those described in \cite{kn:sonner,kn:schalm}.}. We will argue, again using the gauge-gravity duality (but now using a \emph{five-dimensional AdS-Kerr geometry} in the bulk), that a useful generalization of (\ref{ALPHA}) in this situation is
\begin{equation}\label{ALEPH}
{d\hat{\textsf{C}}\over dt}\;<\;{2\over \pi }\,\left(2\;-\;\sqrt{1\;+\;{3\left(\mathcal{A}^2 + \mathcal{B}^2\right)\over L^2}}\right).
\end{equation}
Here the notation is as in the inequality (\ref{ALPHA}), with the addition that $L$ is a certain parameter with units of length (which we need to evaluate explicitly in terms of the physics of the boundary theory), $\mathcal{A}$ and $\mathcal{B}$ (also with units of length) are the two possible parameters describing the ratio of the angular momentum densities to the energy density of the boundary theory (there being two possible simultaneous axes of rotation for a three-sphere)\footnote{In conventional units, (\ref{ALEPH}) is $${d\hat{\textsf{C}}\over dt}\;<\;{2\,c^2\over \pi \hbar}\,\left(2\;-\;\sqrt{1\;+\;{3\left(\mathcal{A}^2 + \mathcal{B}^2\right)\over c^2L^2}}\right).$$}. Of course, the numerical factor is chosen so that (\ref{ALEPH}) reduces to (\ref{ALPHA}) when $\mathcal{A} = \mathcal{B} = 0$.

For the sake of simplicity, and for other reasons to be discussed below, we will set $\mathcal{B} = 0$ until further notice; this simplification does not affect our conclusions in any way (see however \cite{kn:dad} for an example where ``single'' rotation does differ from the general case).

Thus, the rate of growth of the specific complexity of a field theory representing a vortical fluid is still bounded above by a definite value. The value is however a decreasing function of the specific angular momenta, and it can apparently be arbitrarily small (if $\mathcal{A}/L$ can be sufficiently close to unity). That is, \emph{rotation suppresses the growth of complexity}. To put it in a more explicit way: take a boundary fluid with a given, fixed energy density, and ``spin it up'', that is, steadily increase its angular momentum so that $\mathcal{A}$ approaches $L$. Then one can obtain an upper bound on the rate of growth of specific complexity which is much more restrictive than (\ref{ALPHA}).

In which cases is (\ref{ALEPH}) actually significantly more restrictive than (\ref{ALPHA})? We can assess this by using the gauge-gravity duality to translate the problem to one in black hole theory in the bulk.

Assuming that cosmic censorship holds in the AdS context\footnote{As is well known, it is possible to construct seeming counter-examples to censorship for asymptotically AdS systems: see \cite{kn:horsantben,kn:crissant}; but subsequent very remarkable work \cite{kn:horsant} strongly suggests that these can be excluded on the grounds that they conflict with the ``weak gravity conjecture'' \cite{kn:palti}. It is now, therefore, reasonable to assume that censorship does hold for asymptotically AdS black holes, and we will proceed on that basis.}, then we will see that it prevents $\mathcal{A}$ (which, in the bulk interpretation, is the ratio of the black hole angular momentum to its mass) from coming arbitrarily close, not only to the mass (as in the asymptotically flat case) but \emph{also} to $L$ (which, in the bulk, has an interpretation as the asymptotic AdS$_5$ curvature length scale). Consequently the bracketed expression on the right side of (\ref{ALEPH}) cannot be arbitrarily small. It turns out that the minimum possible value for that quantity depends only on the dimensionless physical mass of the black hole, defined as
\begin{equation}\label{BETH}
\mu \equiv {8\ell_{\textsf{B}}^3\mathcal{M}\over \pi L^2},
\end{equation}
where $\ell_{\textsf{B}}$ is the gravitational length scale in the bulk. It is useful to define a function $\mathcal{F}\left(\mathcal{A}/L, \mathcal{B}/L\right)$ by
\begin{equation}\label{BETHA}
\mathcal{F}\left(\mathcal{A}/L, \mathcal{B}/L\right)\;\equiv\; 2\;-\;\sqrt{1\;+\;{3\left(\mathcal{A}^2 + \mathcal{B}^2\right)\over L^2}}.
\end{equation}
Clearly $\mathcal{F}\left(0, 0\right) = 1$, so this function represents the deviation (from (\ref{ALPHA})) induced by rotation. In the case where $\mathcal{B} = 0$, we denote this function by $\mathcal{F}\left(\mathcal{A}/L\right)$.

We will show, assuming censorship, that
\begin{equation}\label{GIMEL}
\mathcal{F}\left(\mathcal{A}/L\right)\;\geq\;{3\left(-1\,+\,\sqrt{9\,+\,8\mu}\right)\over 3\,+\,4\mu \,+\,\sqrt{9\,+\,8\mu}}.
\end{equation}
The expression on the right side is approximately equal to unity when $\mu$ is very small; as $\mu$ increases, it decreases slowly. Thus for a ``small'' black hole (in the sense that $\mu$ is small\footnote{In the non-rotating case, ``smallness'' of the physical mass relative to $L^2$ has also a simple interpretation in terms of the sign of the specific heat of the AdS black hole. In the rotating case, the relation is more complicated, so the reader should not automatically interpret ``smallness'' in that way.}), the standard bound, $d\hat{\textsf{C}}/dt < 2/\pi$, remains approximately valid even if the black hole rotates; but for ``large'' rotating black holes, the rate of growth of specific complexity must be substantially smaller than the standard bound would lead one to expect. To put it another way, this means that (\ref{ALEPH}) is much more restrictive than (\ref{ALPHA}) when the energy density of the boundary field theory is high.

Clearly much remains to be understood regarding the actual physical meaning of (\ref{ALEPH}). Why should rotation suppress the growth of complexity, particularly for high energy densities? Is there any way to quantify this effect in the boundary theory?

We will return to these questions. We propose to approach them, however, by first asking a much more basic question: do we have any actual physical evidence that (\ref{ALPHA}) and (\ref{ALEPH}) are in fact valid? \emph{Can they be tested}? In other words, can they, or some aspect of them, be compared, however indirectly, with relevant experimental data?

We will argue that this is indeed partially possible, in the following very specific sense.

We saw that the inequalities (\ref{ALPHA}) and (\ref{ALEPH}) are derived in two steps. The first involves postulating a relationship between the rate of growth of the specific complexity of a certain field theory and the product of its specific entropy with its temperature (or some generalization of that combination). We do not yet know how to test this hypothesis.

In the second step, one uses the gauge-gravity duality to reduce this part of the problem to that of showing that there is an upper bound on the product of the entropy to mass ratio of an AdS$_5$-Schwarzschild or an AdS$_5$-Reissner-Nordstr\"om black hole with its Hawking temperature. We propose that this second part of the argument, that is, the validity of the gauge-gravity duality in this specific application, \emph{can} be tested, as follows.

One interesting aspect of gauge-gravity duality is concerned with the extent to which the matter described by the boundary field theory resembles the \emph{Quark-Gluon Plasma} (QGP). This is described in great detail in \cite{kn:nat}, and the cautious conclusion is that, when the plasma is very strongly coupled, the resemblance may be close enough to be useful.

If that is the case here, then the second part of our argument above implies that \emph{there must be an upper bound on the product of the specific entropy and the temperature of the actual QGP, and it allows us to compute that upper bound}. That is, this part of the argument implies a specific (if approximate) relationship between ``observed'' quantities in the QGP produced in heavy-ion collisions. (By ``observed'' we mean that some of these quantities must in fact be derived from phenomenological models rather than from direct observations.) We can therefore, while heeding all of the usual warnings regarding the applicability of the AdS/CFT correspondence to actual plasmas, subject part of the argument leading to (\ref{ALPHA}) to ``experimental'' test. It turns out that this is actually a very demanding test, so passing it can be regarded as providing strong support to the whole theory.

In the case of \emph{central} (almost ``head-on'') collisions, which generate negligible angular momentum, this can be done by considering AdS$_5$-Reissner-Nordstr\"om black holes (the electric charge is needed to account for the baryonic chemical potential, which is non-negligible for (some of) the collisions we consider here), and comparing with phenomenological models documented in the literature. In the case of non-central (``peripheral'') collisions, large angular momenta are generated in the QGP, and one should of course use an AdS$_5$-Kerr black hole, with an angular momentum to mass ratio $\mathcal{A} \neq 0$: this parameter corresponds holographically to the ratio of the angular momentum density of the plasma to its energy density.

The comparison with ``experiment'' is more difficult in this case, simply because the relevant data are not yet available. The internal motion of the QGP produced in peripheral heavy-ion experiments manifests itself in the form of polarizations of certain hyperons generated by the collision; from these polarizations one deduces the (average) \emph{vorticity} of the plasma, and consequently the existence of a large (average) angular momentum density. These polarizations have only recently been discovered \cite{kn:STARcoll,kn:STARcoll2} in a remarkable series of observations performed at the RHIC facility \cite{kn:tan}, and many parameters remain unknown or have yet to be modelled. In this case, the black hole inequality underlying (\ref{ALEPH}) can be interpreted as making \emph{predictions} as to what will be found when that analysis is complete; so this part of the argument is again subject to experimental test, in this sense.

We should stress here that the QGP in this case does not of course ``rotate'' in a simple sense: instead it is permeated by very small vortices which propagate through it, so that one speaks of a ``vortical plasma''. When one deduces ``the angular momentum'' of the plasma, or its ``vorticity'', as an explicit numerical quantity (\cite{kn:STARcoll} gives the vorticity as $9\,\pm 1\,\times 10^{21}\,\cdot\,$s$^{-1}$), one is referring to average quantities computed over the volume of the plasma sample (and also over impact energies). That is, no real physical object rotates at this specific angular velocity: this is the angular velocity of a fictitious object which conveniently represents the whole suite of observations of the actual, extremely complex, plasma system. In a holographic model, one seeks a correspondence between the black hole parameter $\mathcal{A}$ and the corresponding parameter for the abstract system which represents the vortical plasma, \emph{not} a detailed model of the rotation of the plasma vortices. (This discussion explains our interest in the special case $\mathcal{B} = 0$: only one angular momentum parameter is needed here.)

We begin by proving the statements regarding AdS$_5$-Reissner-Nordstr\"om and AdS$_5$-Kerr black hole parameters which underlie (\ref{ALPHA}), (\ref{ALEPH}), and (\ref{GIMEL}): this is just a somewhat intricate but elementary computation. We then turn to comparing these statements with phenomenological data, using the gauge-gravity duality in the manner explained above. In the case of central collisions, we find surprisingly good agreement for collision impact energies up to 200 GeV, the highest impact energy studied by the STAR collaboration. Finally we turn to the case of peripheral collisions, and argue that the inequality from which (\ref{ALEPH}) is derived predicts that, at high impact energies, the entropy density of the plasma produced in such collisions may differ significantly from the values characteristic of plasmas generated by central collisions.

For technical reasons to be discussed later, we consider separately the cases of AdS$_5$-Reissner-Nordstr\"om black holes (with AdS$_5$-Schwarzschild as a special case) and AdS$_5$-Kerr black holes. (That is, in the dual theory, we consider plasmas either with large baryonic chemical potential and negligible vorticity, or the reverse. The possibility that \emph{both} vorticity and baryonic chemical potential might be large will be discussed later.)

\addtocounter{section}{1}
\section* {\large{\textsf{2. AdS$_5$-Reissner-Nordstr\"om Black Holes}}}
Before we begin, we stress that, throughout this work, we are exclusively concerned with \emph{five-dimensional} asymptotically AdS black holes. These can behave in quite a different way from their four-dimensional counterparts, and we warn the reader that many of the statements made below do not apply in other numbers of spacetime dimensions \cite{kn:reall}.

An AdS$_5$-Reissner-Nordstr\"om black hole with a spherical event horizon has the metric
\begin{flalign}\label{ONE}
g(\m{RNAdS}_5)\;=\;&-\,\left({r^2\over L^2}\,+\,1\,-\,{2M\over r^2}\,+\,{Q^2\over 4\pi r^4}\right)\m{d}t^2\,+{\m{d}r^2\over {r^2\over L^2}\,+\,1\,-\,{2M\over r^2}\,+\,{Q^2\over 4\pi r^4}}\\ \notag \,\,\,\,&\,+\,r^2\left(\m{d}\theta^2 \,+\, \sin^2\theta\,\m{d}\phi^2\,+\,\cos^2\theta\,\m{d}\psi^2\right).
\end{flalign}
Here $t$ and $r$ are as usual, and ($\theta,\,\phi,\,\psi$) are (Hopf) coordinates on a three-dimensional sphere. The geometric parameters $M$ and $Q$ are related to the physical mass and charge of the black hole, $\mathcal{M}$ and $\mathcal{Q}$, by
\begin{equation}\label{TWO}
\mathcal{M}\;=\;{3\pi M\over 4\ell^3_{\textsf{B}}},\;\;\;\;\;\mathcal{Q}\;=\;{\sqrt{3}\pi Q\over 2\ell^{3/2}_{\textsf{B}}},
\end{equation}
where $\ell_{\textsf{B}}$ is the gravitational length scale in the bulk. Holographically, $\mathcal{M}$ is related to the energy density of the boundary field theory, and $\mathcal{Q}$ is proportional to its baryonic chemical potential.

The Hawking temperature of this black hole is given by
\begin{equation}\label{THREE}
4\pi T\;=\;{2\over r_H}\,+\,{4r_H\over L^2}\,-\,{Q^2\over 2\pi r_H^5},
\end{equation}
where $r_H$ locates the (outer) event horizon:
\begin{equation}\label{FOUR}
{r_H^2\over L^2}\,+\,1\,-\,{2M\over r_H^2}\,+\,{Q^2\over 4\pi r_H^4}\;=\;0.
\end{equation}
The black hole entropy is
\begin{equation}\label{FIVE}
S\;=\;{\pi^2r_H^3\over 2\ell^3_{\textsf{B}}}.
\end{equation}

Combining these relations, we find, after some manipulation and simplification,
\begin{equation}\label{SIX}
{ST\over \mathcal{M}}\;=\;{2\over 3}\,\left[{1 + {2r_H^2\over L^2} - {Q^2\over 4\pi r_H^4}\over 1 + {r_H^2\over L^2} + {Q^2\over 4\pi r_H^4}}\right].
\end{equation}
As we will rehearse in detail in the succeeding Section, the rate of growth of the specific complexity is bounded \cite{kn:suss2,kn:jacob} by a fixed multiple of $ST/\mathcal{M}$, so this is the quantity we need to understand here. We note an important and very useful property of this quantity: it does not depend on $\ell_{\textsf{B}}$, which we do not know. In fact we regard this as an indication that $ST/\mathcal{M}$ is an interesting quantity to consider, independently of its role in computations of the rates of growth of specific complexities.

The right side of (\ref{SIX}) behaves in two different ways, according to whether $Q = 0$ or not.

When $Q = 0$, one finds that the right side of (\ref{SIX}), regarded as a function of $r_H$, is a monotonically increasing function which is bounded \emph{both below and above}: the lower limit is obtained as $r_H \rightarrow 0$ (which in this case implies $\mathcal{M} \rightarrow 0$), the upper limit as $r_H \rightarrow \infty$:
\begin{equation}\label{SEVEN}
{2\over 3}\;<\;{ST\over \mathcal{M}}\;<\;{4\over 3}\;\;\;\; \left(\mathcal{A} = \mathcal{Q} = 0\right).
\end{equation}
With a suitable choice \cite{kn:suss3} of the constant relating the growth of specific complexity to $ST/\mathcal{M}$, this is (\ref{ALPHA}), but with the additional information that there is a \emph{lower} bound, equal to one half of the upper bound.

This lower bound depends, unfortunately, on $Q$ (or $\mathcal{Q}$) being precisely zero. For if $Q \neq 0$, one finds that (no matter how small $Q$ may be) the lower bound is simply zero (corresponding now to an extremal black hole with zero temperature). The function continues to be a monotonically increasing one, and it continues to be (asymptotically) bounded above: in fact, \emph{the upper bound is completely independent} of $Q$, which means that it continues to be equal to $4/3$:
\begin{equation}\label{EIGHT}
0\;<\;{ST\over \mathcal{M}}\;<\;{4\over 3}\;\;\;\; \left(\mathcal{A} = 0,\;\; \mathcal{Q} \neq 0\right).
\end{equation}
With a suitable identification of the constant relating $ST/\mathcal{M}$ to a bound on $d\hat{\textsf{C}}/dt$, this is (\ref{ALPHA}).

Notice that the bound in (\ref{EIGHT}) is optimal, in the sense that, given any $c < 4/3$, one can find a black hole with parameters such that $ST/\mathcal{M} > c\,$; indeed, one can find many. To see this, prescribe arbitrary non-zero values of $Q$ and $L$; since the right side of (\ref{SIX}) takes (for all non-zero $Q$ and $L$) all values between zero and $4/3$, one can set it equal to some number strictly between $c$ and $4/3$ and solve for $r_H$. Then (\ref{FOUR}) gives the appropriate value of $M$.

Thus we see that, in the five-dimensional case, there is an upper bound on $ST/\mathcal{M}$, the \emph{same} for all AdS$_5$-Schwarzschild and AdS$_5$-Reissner-Nordstr\"om black holes. This means that we have not succeeded in improving the bound on the rate of growth of the specific complexity in the charged case; for that, one should consult \cite{kn:suss3,kn:zhong}. Of course, this does not show that (\ref{ALPHA}) is incorrect in the charged case, only that it is not optimal. As we will see, however, there is much of interest to be learned even from this simple analysis.

\addtocounter{section}{1}
\section* {\large{\textsf{3. AdS$_5$-Kerr Black Holes}}}
The AdS$_5$-Kerr metric (with no electric or magnetic charge) \cite{kn:hawk,kn:cognola,kn:gibperry} is, in the most general case, characterized by a mass parameter $M$ and by two rotation parameters $(a,b)$ (both with units of length). For reasons we have explained, we focus here on the special case with $b = 0$. The metric is then
\begin{flalign}\label{A}
g\left(\m{AdSK}_5^{(a,0)}\right) = &- {\Delta_r \over \rho^2}\Bigg[\,\m{d}t \; - \; {a \over \Xi}\m{sin}^2\theta \,\m{d}\phi\Bigg]^2\;+\;{\rho^2 \over \Delta_r}\m{d}r^2\;+\;{\rho^2 \over \Delta_{\theta}}\m{d}\theta^2 \\ \notag \,\,\,\,&+\;{\m{sin}^2\theta \,\Delta_{\theta} \over \rho^2}\Bigg[a\,\m{d}t \; - \;{r^2\,+\,a^2 \over \Xi}\,\m{d}\phi\Bigg]^2 \;+\;r^2\cos^2\theta \,\m{d}\psi^2 ,
\end{flalign}
where
\begin{eqnarray}\label{B}
\rho^2& = & r^2\;+\;a^2\m{cos}^2\theta, \nonumber\\
\Delta_r & = & (r^2+a^2)\Big(1 + {r^2\over L^2}\Big) - 2M,\nonumber\\
\Delta_{\theta}& = & 1 - {a^2\over L^2} \, \m{cos}^2\theta, \nonumber\\
\Xi & = & 1 - {a^2\over L^2}.
\end{eqnarray}
Here $L$ is again the asymptotic AdS curvature length scale, and the coordinates are as in the preceding Section.

If $\mathcal{M}$ denotes the \emph{physical} mass (that is, the mass that appears in the first law of thermodynamics \cite{kn:gibperry} for these black holes) and $\mathcal{J}$ is the physical angular momentum, then it is shown in \cite{kn:gibperry} that
\begin{equation}\label{C}
\mathcal{M}\;=\;{\pi M \left(2 + \Xi\right)\over 4\,\ell_{\textsf{B}}^3\,\Xi^2}, \;\;\;\;\;\mathcal{J}\;=\;{\pi M a\over 2\,\ell_{\textsf{B}}^3\,\Xi^2},
\end{equation}
where $\ell_{\textsf{B}}$ is the bulk gravitational length scale, as before. (As mentioned earlier, in natural units $\mathcal{M}$ has units of 1/length; $\mathcal{J}$ is dimensionless.) The angular momentum to (physical) mass ratio $\mathcal{A}$ (units of length) is therefore given by
\begin{equation}\label{D}
\mathcal{A}\;=\;{2 a \over 2 + \Xi}\;=\;{2 a \over 3 - \left(a^2/L^2\right)}.
\end{equation}

In this work, we are interested in a comparison of the boundary system here with the QGP. We imagine the latter as initially having essentially zero angular momentum; it acquires its angular momentum as a consequence of processes initiated by a collision of heavy ions. Therefore we suppose that our black hole is obtained by steadily increasing $\mathcal{A}$ from zero: we imagine that we are continuously ``spinning up'' the black hole from an initial non-rotating state. (That is, we ignore the possibility that the angular momentum might be ``primordial''.) It is clear from the definition of $\Xi$ and from the form of the relations (\ref{C}) that, in this case, we must require $a < L$. It is straightforward to show that this implies
\begin{equation}\label{DD}
\mathcal{A}\;<\;L,
\end{equation}
and we will hold to this henceforth; it will be important in Section 5 below.

The Hawking temperature of the AdS$_5$-Kerr black hole is given \cite{kn:gibperry} by
\begin{equation}\label{E}
T\;=\;{r_H\left(1 + {r_H^2\over L^2}\right)\over 2\pi \left(r_H^2 + a^2\right)} + {r_H\over 2\pi L^2},
\end{equation}
where $r_H$ denotes the horizon ``radius'' (the largest root of $\Delta_r$), and the entropy is
\begin{equation}\label{F}
S\;=\;{\pi^2\left(r_H^2 + a^2\right)r_H\over 2\ell_{\textsf{B}}^3\Xi}.
\end{equation}
The ratio $S/\mathcal{M}$ is
\begin{equation}\label{G}
{S\over \mathcal{M}}\;=\;{2\pi r_H\left(r_H^2+a^2\right)\Xi\over M\left(2 + \Xi\right)}.
\end{equation}

Using the definition of $r_H$, we have $2M = \left(r_H^2 + a^2\right)\left(1 + {r_H^2\over L^2}\right)$, and inserting this into (\ref{G}) we have
\begin{equation}\label{H}
{S\over \mathcal{M}}\;=\;{4\pi r_H\Xi\over \left(2 + \Xi\right)\left(1 + {r_H^2\over L^2}\right)}.
\end{equation}

Combining this with equation (\ref{E}), we have
\begin{equation}\label{I}
{ST\over \mathcal{M}}\;=\;{2\,\Xi\over 2 + \Xi}\,r_H^2\,\left({1\over r_H^2 + a^2}\,+\,{1\over r_H^2 + L^2}\right).
\end{equation}
Again we regard this as a function of $r_H$; once again, it increases monotonically towards an asymptotic value. The case $\mathcal{A} = a = 0$ was considered in the preceding Section. When $\mathcal{A} \neq 0$, there is no lower bound other than zero (corresponding to $r_H \rightarrow 0$, which is the extremal case, but now with $\mathcal{M} \neq 0$; see below), but we have, considering the limit $r_H \rightarrow \infty$,
\begin{equation}\label{K}
{ST\over \mathcal{M}}\;<\;{4\,\Xi\over 2 + \Xi}.
\end{equation}
As in the preceding Section, this cannot be improved; given any $c$ (strictly) between zero and $4\,\Xi/(2 + \Xi)$, there exist black holes with any given $a$ and $L$, but with $ST/\mathcal{M} > c$.

Now $\Xi$ is defined by $a/L$, which is related to $\mathcal{A}/L$ by equation (\ref{D}); so $\Xi$ can be expressed in terms of $\mathcal{A}/L$, as follows:
\begin{equation}\label{L}
\Xi\;=\;{\sqrt{1\,+\,{3\mathcal{A}^2\over L^2}}\;-\;1\;-\;{\mathcal{A}^2\over L^2}\over {\mathcal{A}^2\over 2L^2}}.
\end{equation}
Substituting this into (\ref{K}), we have, after a lengthy simplification,
\begin{equation}\label{M}
{ST\over \mathcal{M}}\;<\;{4\, \over 3 }\left(2\;-\;\sqrt{1\;+\;{3\mathcal{A}^2\over L^2}}\right)\;\;\;\;\left(\mathcal{Q} = 0,\;\mathcal{A} \neq 0\right).
\end{equation}

This inequality was derived in the case where the second possible rotation parameter, $b$ (which defines a second physical angular momentum per unit mass, $\mathcal{B}$), has been set equal to zero. However, in the fully general case (see \cite{kn:gibperry}), the metric and the parameter correspondences are completely symmetric with respect to the formal transformation $a \leftrightarrow b$, combined with a simple coordinate change, $\theta \leftrightarrow \pi/2 - \theta$. It follows that the extension of (\ref{M}) to this case is
\begin{equation}\label{N}
{ST\over \mathcal{M}}\;<\;{4\, \over 3 }\left(2\;-\;\sqrt{1\;+\;{3\left(\mathcal{A}^2 + \mathcal{B}^2\right)\over L^2}}\right)\;\;\;\;\left(\mathcal{Q} = 0,\;\mathcal{A},\,\mathcal{B} \neq 0\right).
\end{equation}
We see that the inclusion of the second rotation parameter only strengthens the effect of the first. We also see that, unlike $Q$, $\mathcal{A}$ \emph{does} affect the bound on this quantity.

This inequality describes the situation in the bulk. Now let us consider the interpretation of these quantities and this inequality in the dual system on the boundary.

It is thought \cite{kn:suss2,kn:jacob} that the rate of growth of complexity can be bounded as follows. The entropy $S_{\partial}$ of the boundary system (more precisely, $S_{\partial}/k_B$, where $k_B$ is the Boltzmann constant) can be regarded as counting the maximal possible number of relevant degrees of freedom, while $\hbar/\left(k_BT_{\partial}^*\right)$, where $T_{\partial}^*$ is the temperature of the boundary matter measured in a frame that rotates with that matter, gives the appropriate time scale, again measured in the rotating frame. Therefore the maximal rate, measured by this time coordinate (let us call it $t^*$), at which the complexity of this system grows (that is, the maximal ``rate at which gates are executed''), $d\textsf{C}/dt^*$, should be\footnote{As in the AdS$_5$-Reissner-Nordstr\"om case, this part of the argument can \emph{also} be strengthened for AdS$_5$-Kerr black holes: see again \cite{kn:suss3,kn:zhong}. Thus, again, we do not claim that the bound we obtain is optimal.} bounded by some fixed dimensionless multiple of $S_{\partial}T_{\partial}^*/\hbar$.

Clearly $t^*$ differs from the time $t$ measured by a non-rotating observer at infinity, but on the other hand $T_{\partial}^*$ also differs from the temperature $T_{\partial}$ measured by that observer. As is explained in \cite{kn:jacob}, the two effects cancel: $T_{\partial}^*dt^* = T_{\partial}dt$, so we find that $d\textsf{C}/dt$ should be bounded by a fixed dimensionless multiple of $S_{\partial}T_{\partial}/\hbar$.

We now replace $\textsf{C}$ by the specific complexity $\hat{\textsf{C}}$, the entropy $S_{\partial}$ by the specific entropy, and appeal to holography to translate the resulting statement to the bulk. We equate the boundary specific entropy to the black hole quantity $S/\mathcal{M}$, $T_{\partial}$ to the Hawking temperature $T$ of the black hole, and the ratios of the boundary theory angular momentum densities to the energy density to the black hole angular momentum parameters $\mathcal{A}$ and $\mathcal{B}$. We then see that (\ref{N}) gives the new bound, (\ref{ALEPH}) (up to a dimensionless factor which is fixed in the manner explained earlier).

Henceforth we resume setting $\mathcal{B} = 0$, partly to avoid the great complexity of the general case, partly because $\mathcal{B}$ has no clear physical role in the dual theory, as discussed earlier.

The extent to which (\ref{ALEPH}) can differ from (\ref{ALPHA}) is determined by considering cosmic censorship. To see this, we begin with the observation that censorship in this case requires that $\Delta_r = 0$ (see the equations (\ref{B})) should have a real, positive solution. The condition for that is apparently very simple:
\begin{equation}\label{P}
a^2\;< \;2M.
\end{equation}
In terms of the physical parameters $\mathcal{M}$ and $\mathcal{A}$, it is considerably less simple:
\begin{equation}\label{Q}
L^2\left(1 - \Xi\right)\;< \; {8\ell_{\textsf{B}}^3\mathcal{M}\Xi^2\over \pi \left(2 + \Xi\right)},
\end{equation}
which can be re-written as
\begin{equation}\label{R}
\left(\mu + 1\right)\Xi^2 + \Xi - 2 \;> \; 0;
\end{equation}
here $\mu \equiv 8\ell_{\textsf{B}}^3\mathcal{M}/(\pi L^2)$ is the dimensionless physical mass mentioned earlier, and we are thinking of $\Xi$ as a proxy for $\mathcal{A}/L$ (see equation (\ref{L}) above). The quadratic in $\Xi$ on the left has one positive real root, $\Xi_+$, which can be readily expressed in terms of $\mu$; for (\ref{R}) to be satisfied we must have $\Xi > \Xi_+$. Through equation (\ref{L}), this puts an upper bound on $\mathcal{A}/L$:
\begin{equation}\label{S}
{\mathcal{A}\over L} \;<\;{2\sqrt{1 - \Xi_+}\over 2 + \Xi_+} \;<\;1,
\end{equation}
the inequalities being strict. The right side of this relation can be regarded as a certain (increasing) function of $\mu$. Thus cosmic censorship, assuming that it holds, prevents $\mathcal{A}$ from being too large relative to the dimensionless mass (or, more precisely, too large relative to a quantity which acts as proxy for the dimensionless mass). However, we see that it also prevents $\mathcal{A}$ from approaching $L$ too closely.

Inserting (\ref{S}) into the right side of (\ref{K}) and expressing $\Xi_+$ in terms of $\mu$, one obtains, after some further simplifications, the inequality (\ref{GIMEL}) given in the preceding section: thus we can specify how close to $L$ we can allow $\mathcal{A}$ to become. As claimed in Section 1, $\mathcal{A}$ must be much smaller than $L$ when $\mu$ is small, but it can approach $L$ more closely when $\mu$ is large. Thus our improvement of the upper bound is most relevant for ``large'' black holes.

We saw above (see (\ref{SEVEN})) that, when $\mathcal{A} = 0$, there is actually a \emph{lower} bound on $ST/\mathcal{M}$. This is quite reasonable from the point of view of a putative ``second law of complexity''. Unfortunately, however, this lower bound apparently disappears as soon as the black hole either acquires a charge or rotates; as with any charged or rotating black hole, by increasing the charge or the angular momentum towards the extremal value, one can force $ST$ to be arbitrarily small.

However, one should note that the uncharged, rotating black holes we are considering in this Section have an unusual property: as extremality is approached, the area of the event horizon also becomes arbitrarily small. Thus, a near-extremal AdS$_5$-Kerr black hole can conceal an arbitrarily large mass within a surface of arbitrarily small area. This is simply because, in five dimensions \cite{kn:reall}, unlike in four, the condition for extremality forces $r_H$ to vanish; thus a near-extremal black hole has a very small event horizon in this case. (Actually, a truly extremal black hole in this case (with $a^2 = 2M$) does not exist: one has instead an (effectively) naked ring singularity. See \cite{kn:emparanmyers}.)

It may well be that such objects, with large masses and arbitrarily small event horizon areas, are unphysical: see for example \cite{kn:marolf,kn:ong} for discussions of the issues. (Similarly, sufficiently near-extremal AdS$_5$-Reissner-Nordstr\"om black holes may not be physical; this is known to be the case for charged asymptotically AdS black holes with flat event horizons \cite{kn:bounding}.) It may be possible to use this to establish a lower bound on $ST$ given the mass, and so to forge a link with the second law of complexity. Thus, while the lower bound in (\ref{SEVEN}) is not ``stable'', in the sense that it disappears for any non-zero charge or angular momentum, it should not be dismissed altogether: it offers a hint that one may ultimately be able to prove a ``stable'' version.

Finally, one might ask: what happens when \emph{both} the charge on the black hole, and its angular momentum, are non-zero? The obvious way to approach this would be to use a five-dimensional version of the AdS-Kerr-Newman geometry in the bulk. Unfortunately, however, this metric is not yet known, even in the asymptotically flat case \cite{kn:emparan}. However, it seems very likely that, just as the upper bound on $ST/\mathcal{M}$ proved to be independent of the charge for AdS$_5$-Reissner-Nordstr\"om black holes, the same is true here: that is, we do not expect any modification to (\ref{N}) or (\ref{ALEPH}) to result from including electric charge. (Some support for this statement can be derived from examining the approximate (slowly rotating, asymptotically flat) five-dimensional Kerr-Newman metric given in \cite{kn:aliev}, in which the charge appears in a way that suggests that it would not affect the upper bound.) We will return to this point in Section 5 below.

We now turn to the question of testing (\ref{ALPHA}) and (\ref{ALEPH}) ``experimentally''.

\addtocounter{section}{1}
\section* {\large{\textsf{4. Testing the Bound: Central Collisions}}}
Central collisions of heavy ions have been extensively studied, and good phenomenological models of the resulting plasmas are available: we will use \cite{kn:sahoo}. Thus one has reasonably reliable estimates of the entropy density $s$, its energy density $\varepsilon$, and the temperature $T$ (though this last is particularly uncertain \cite{kn:bus}).

For \emph{central} collisions, the angular momentum density of the plasma is negligible, but the baryonic chemical potential is not (except at the highest impact energies), and so the dual geometry is that of an AdS$_5$-Reissner-Nordstr\"om black hole: see the inequality (\ref{EIGHT}), above.

Holography maps the bulk quantity $S/\mathcal{M}$ to $s/\varepsilon$, and the Hawking temperature to the temperature of the plasma. The inequality (\ref{EIGHT}) now takes the form
\begin{equation}\label{T}
{s\,T\over \varepsilon}\;<\;{4\over 3}.
\end{equation}
Estimates for all of the quantities appearing in this relation can be found in \cite{kn:sahoo}, and so we can indeed check our results (in this case) against ``experiment''.

Before proceeding to that, we stress that (\ref{T}) is a stringent test of the theory, in the sense that not all of the components are of order unity in natural units: for example, from \cite{kn:sahoo} one finds that, for central collisions at impact energy 200 GeV, the entropy density $s$ is predicted to be about 13.5 (fm)$^{-3}$. Thus $sT/\varepsilon$ might easily have been larger than a quantity of order unity.

The STAR collaboration \cite{kn:STAR} reports data corresponding to impact energies ranging from 7.7 to 200 GeV, and \cite{kn:sahoo} gives values accordingly. In Figure 1 we show the value of the combination $sT/\varepsilon$ for these energies, and also the proposed upper bound for comparison.

\begin{figure}[!h]
\centering
\includegraphics[width=1.1\textwidth]{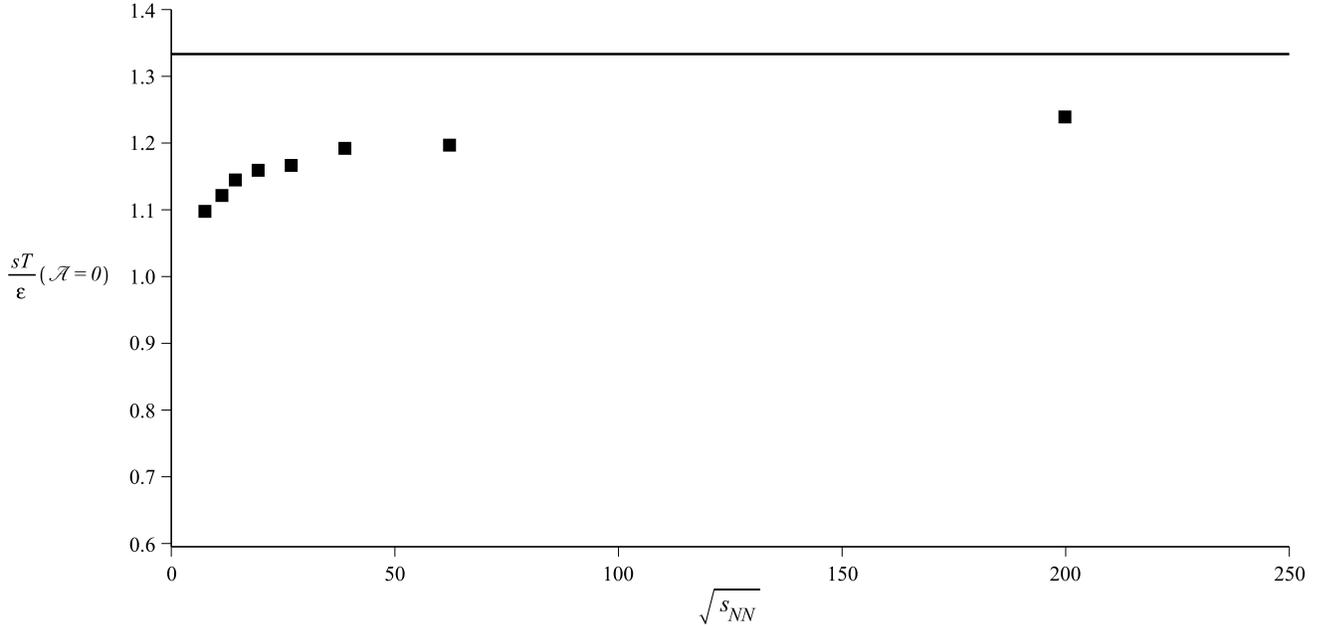}
\caption{Phenomenological estimates, after \cite{kn:sahoo}, for $sT/\varepsilon$ (which is dimensionless in natural units), corresponding to central collisions at $\sqrt{s_{\m{NN}}} = 7.7,\, 11.5,\, 14.5,\, 19.6,\, 27,\, 39,\, 62.4,\, 200$ GeV. The horizontal line is 4/3, the proposed upper bound.}
\end{figure}
There are several points to be made regarding this figure. The first, of course, is that (\ref{T}) is satisfied in every case: this part of the argument leading to the complexity bound (\ref{ALPHA}) is fully in agreement with these data. To that extent, we have evidence in favour of that bound. Secondly, there is a clear upward trend with increasing impact energy; the bound is most nearly attained by collisions at the highest impact energies. This is as expected holographically, for in (\ref{SIX}) the upper bound is approached when the event horizon is large, and this corresponds to black holes with large masses or energies.

Of course, one should not expect any great precision from a simple holographic model, and the surprisingly good agreement of theory with data represented in Figure 1 need not be taken too literally. Nevertheless, it seems reasonable to assert that some upper bound on $sT/\varepsilon$ does exist in this regime of impact energies, and that the upper bound is probably not very far from 4/3.

Indeed, the limitations of holographic models of the QGP should also be borne in mind regarding another aspect of Figure 1. We may ask: what happens at still higher impact energies? The answer is that we should not expect the model to deal with this. One does not expect the gauge-gravity duality to apply to the QGP universally, only to those examples which are indeed \emph{strongly coupled}. At high temperatures, the plasma is not expected to be strongly coupled, so (\ref{T}) ceases to be relevant.

One simple way to assess this is to examine the baryonic chemical potential: we can expect it to be negligible (relative to the temperature) for very high-energy collisions, and these are the ones for which strong coupling fails. At 200 GeV, the baryonic chemical potential is around \cite{kn:phobos,kn:sahoo} 27 MeV, already quite small relative to the temperature, which \cite{kn:sahoo} is approximately 190 MeV. At higher impact energies, the baryonic chemical potential continues to fall, and the temperature continues to rise. It is in fact reasonable to suppose that, beyond an impact energy of about 300 GeV, for which the baryonic chemical potential of the plasma is quite negligible relative to the temperature, the QGP is too weakly coupled for a gauge-gravity treatment to be appropriate. We expect (\ref{T}) to continue to hold up to about that impact energy.

In summary, we can claim that, to the extent that the actual QGP produced in central collisions is approximated by the boundary field theory dual to an AdS$_5$-Reissner-Nordstr\"om black hole, the phenomenological data are in agreement with (\ref{EIGHT}). Assuming the existence of a bound on the rate of specific complexity growth by a multiple of the product of the specific entropy with the temperature, we can say that, in this peculiar sense, the complexity bound (\ref{ALPHA}) has ``experimental'' support.

\addtocounter{section}{1}
\section* {\large{\textsf{5. Testing the Bound: Peripheral Collisions}}}
In this section we attempt to repeat the procedure in the previous Section, but now using (\ref{M}). Clearly the most interesting case is the one with largest $\mathcal{A}$; holographically, this means that we wish to consider the QGP with large values of the angular momentum density. (Recall that (\ref{M}) is derived under the assumption that the baryonic chemical potential is negligible, which is \emph{not} the case for most of the impact energies we consider. As explained earlier, we do not expect this to matter; but also see the end of this section for a further discussion.)

The QGP acquires an angular momentum in peripheral collisions; there are good phenomenological models for this; we use \cite{kn:jiang}. There it is shown that the angular momentum depends in an interesting way on both the impact energy and the centrality (that is, essentially, the impact parameter) of the collision. The angular momentum is of course zero for exactly central collisions, but it rises extremely rapidly with increasing centrality, reaching (for collisions of gold nuclei, as used in the RHIC experiments) a maximum at around $7\%$, and then decreasing. From this, we can compute \cite{kn:93} the angular momentum density $\alpha$ for a given centrality and impact energy; it attains its maximum at about $17\%$, larger than for the angular momentum itself (because the relevant volume also depends on the centrality: see \cite{kn:95}), but still quite small.

For these low centralities, we do not expect the energy density or the temperature to differ significantly from the central case; so using \cite{kn:sahoo}, we can compute, for small impact parameters and for a range of impact energies, the ratio $\alpha/\varepsilon$, where $\varepsilon$ is, as before, the energy density. The black hole parameter $\mathcal{A}$ is holographically dual to $\alpha/\varepsilon$, so now we can compute this quantity from the phenomenological models. As explained above, $\alpha$ is largest for impacts with a specific centrality, and we concentrate on those collisions. We have computed $\mathcal{A}$ in this manner for the same range of impact energies as in the preceding Section.

In order to proceed, we must first deal with the fact that there is an important difference between (\ref{ALPHA}) and (\ref{ALEPH}) (and between (\ref{EIGHT}) and (\ref{M})), namely, that (\ref{ALEPH}) and (\ref{M}) involve $L$. We need to estimate this parameter. We already in fact have a strong hint for this: the inequality (\ref{DD}), which asserts that $L$ must be larger than $\mathcal{A}$. There are various possible interpretations of this statement, but the simplest one is to take it to mean that $L$ must be larger than the \emph{largest possible} value of $\mathcal{A}$.

In principle, $\mathcal{A}$ can of course be arbitrarily large, so we have to interpret this statement as referring to the largest value that $\mathcal{A}$ can take for \emph{strongly coupled} plasmas, the only ones, as mentioned above, for which the gauge-gravity duality can be expected to yield useful results. Now in fact $\mathcal{A}$ (computed at the centrality at which it is maximal for a fixed impact energy) increases almost linearly with the impact energy: see Figure 2. We can therefore extrapolate to find the value of $\mathcal{A}$ corresponding to impact energies beyond which strong coupling ceases to hold. On the basis that the baryonic chemical potential is essentially zero beyond impact energy 300 GeV or so, we take from Figure 2 that the maximal value (in this sense) of $\mathcal{A}$ is about 113 fm, and this gives us $L$.
\begin{figure}[!h]
\centering
\includegraphics[width=1\textwidth]{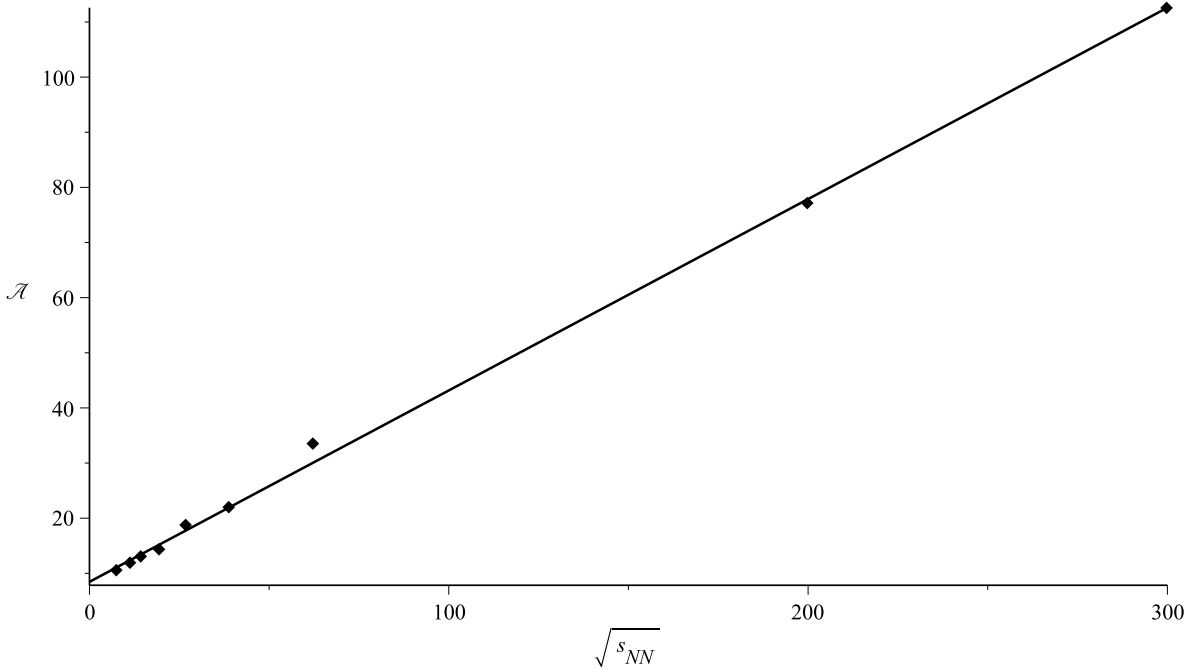}
\caption{Phenomenological estimates, after \cite{kn:sahoo} and \cite{kn:jiang}, for $\mathcal{A}$, corresponding to collisions at centrality $\approx 17\%$ and impact energies $\sqrt{s_{\m{NN}}} = 7.7,\, 11.5,\, 14.5,\, 19.6,\, 27,\, 39,\, 62.4,\, 200$ GeV. Units on the vertical axis are fm.}
\end{figure}

As an (important) aside, we note that the volume of the spatial sections at infinity is given by \cite{kn:93}
\begin{equation}\label{U}
V(L, \mathcal{A})\;=\;2\pi^2L^3\left[{2L^2\over a^2}\,\left({1\over \sqrt{\Xi}}\,-\,1\right)\right].
\end{equation}
For collisions of gold nuclei at a centrality that maximizes the angular momentum density, and at impact energy 200 GeV, one finds that $\mathcal{A} \approx 77$ fm. Using this and $L \approx 113$ fm, one can compute (equation (\ref{D})) $a \approx 90.6$ fm, and then the volume of the spatial section proves to be $\approx 5.95 \times 10^7$ (fm)$^3$. Compared to the volume of the overlap zone for such a collision (no more than about 100 (fm)$^3$), this is essentially infinite. Thus we see that the compactness of the spatial sections at infinity is not a problem here.

The version of (\ref{M}) relevant to the boundary field theory is
\begin{equation}\label{V}
{sT\over \varepsilon}\;<\;{4\, \over 3 }\left(2\;-\;\sqrt{1\;+\;{3\mathcal{A}^2\over L^2}}\right),
\end{equation}
and we are now able to compute the right side of this inequality. Unfortunately, however, we cannot now compare the right side with the left, because we do not have phenomenological models for $s$, $T$, or $\varepsilon$ when large vorticities are present. Thus, we are unable to confirm the upper bound in the manner of the preceding Section. Instead, we can use the bound to \emph{predict} the effect of vorticity on $sT/\varepsilon$. Specifically, we propose the following approach.

We have already argued that, for relatively small centralities, we do not expect $T$ or $\varepsilon$ to vary significantly from their values in central collisions. This is much less clear in the case of $s$, as we will discuss: so let us proceed by regarding (\ref{V}) as giving an upper bound on $s$:
\begin{equation}\label{W}
s\;<\;{4\,\varepsilon \over 3\,T}\,\left(2\;-\;\sqrt{1\;+\;{3\mathcal{A}^2\over L^2}}\right);
\end{equation}
with our assumptions, the right side can now be evaluated. We will denote it by $s_P^+$, to indicate that it is indeed an upper bound, and that it applies to peripheral collisions.

We have computed $s_P^+$, using \cite{kn:sahoo} and \cite{kn:jiang}, for all of the impact energies we considered in the preceding Section. In Figure 3 we show the results.
\begin{figure}[!h]
\centering
\includegraphics[width=1.1\textwidth]{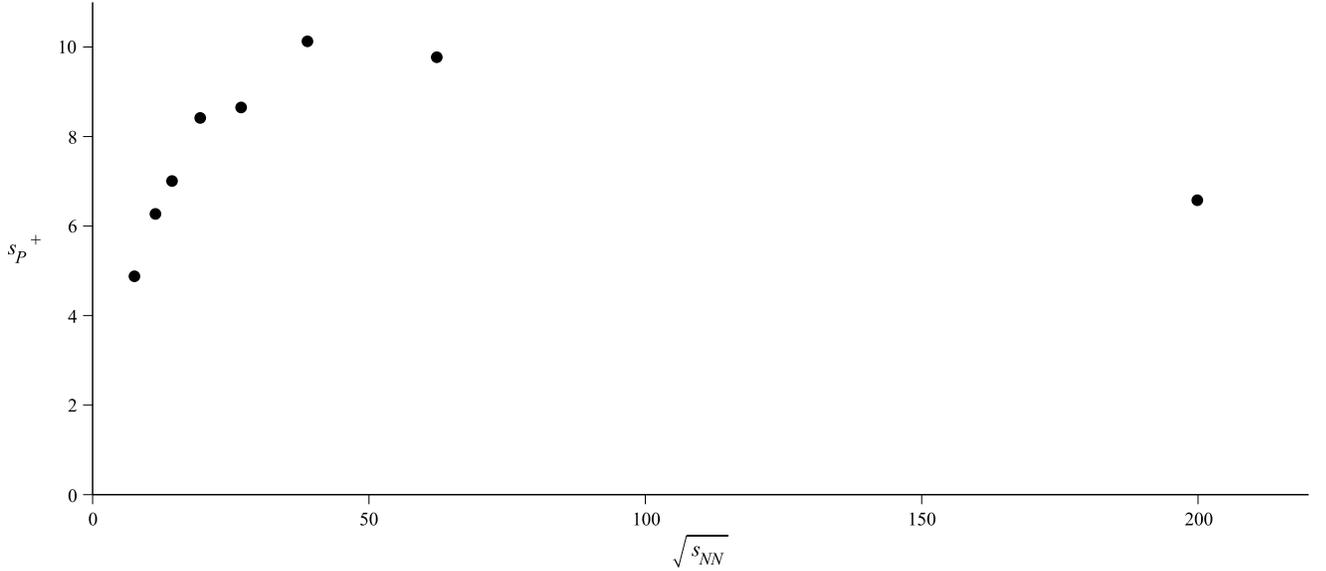}
\caption{Theoretical upper bounds, $s_P^+$, on the entropy densities of the plasmas produced in peripheral collisions at $17\%$ centrality and at $\sqrt{s_{\m{NN}}} = 7.7,\,11.5,\, 14.5,\, 19.6,\, 27,\, 39,\, 62.4,\, 200$ GeV. Units on the vertical axis are (fm)$^{-3}$.}
\end{figure}
We see that, at low impact energies, the bound increases sharply with impact energy; but that, beyond 39 GeV, it decreases.

In order to clarify the meaning of Figure 3, we show in Figure 4 the actual entropy densities $s_C$ (not upper bounds) in the corresponding central collisions. Of course, these need not necessarily be smaller than $s_P^+$, but they allow us to see the effects of angular momentum more clearly.

\begin{figure}[!h]
\centering
\includegraphics[width=1.1\textwidth]{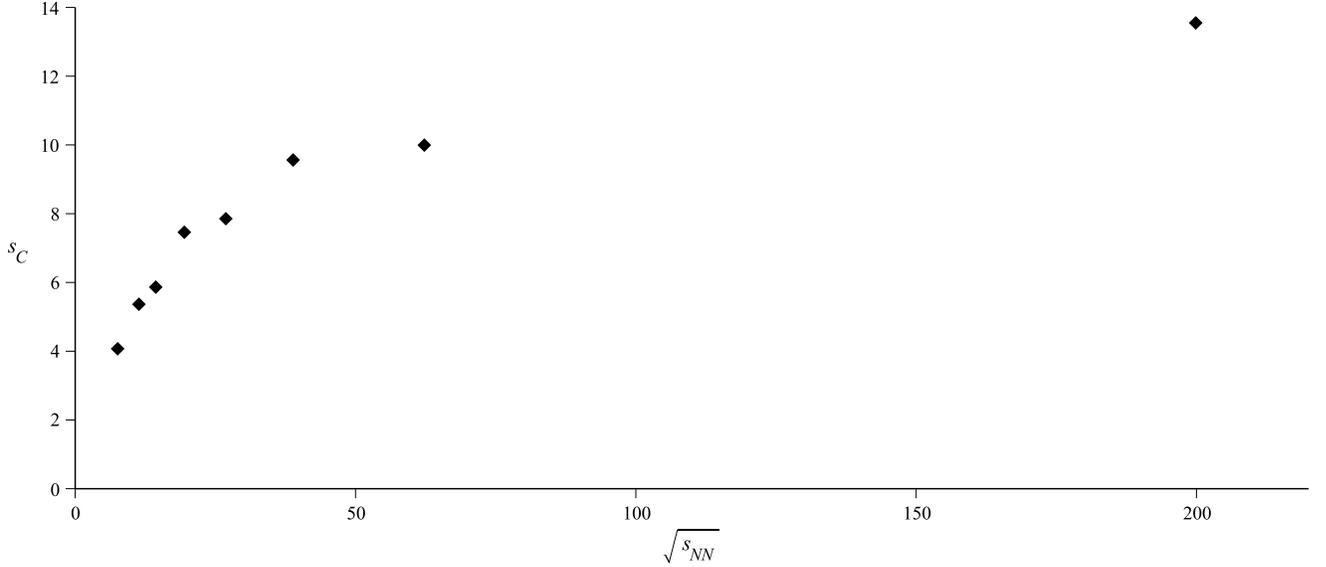}
\caption{The entropy densities (from \cite{kn:sahoo}), $s_C$, of the plasmas produced in central collisions at $\sqrt{s_{\m{NN}}} = 7.7,\,11.5,\, 14.5,\, 19.6,\, 27,\, 39,\, 62.4,\, 200$ GeV. Units on the vertical axis are (fm)$^{-3}$.}
\end{figure}

Up to 39 GeV, the points in Figure 4 are all lower than the corresponding points in Figure 3. In order to ease the comparison of Figure 3 with Figure 4, we have constructed a spline interpolation of the points in Figure 3, and superimposed it on Figure 4; this is shown in Figure 5.

\begin{figure}[!h]
\centering
\includegraphics[width=1.1\textwidth]{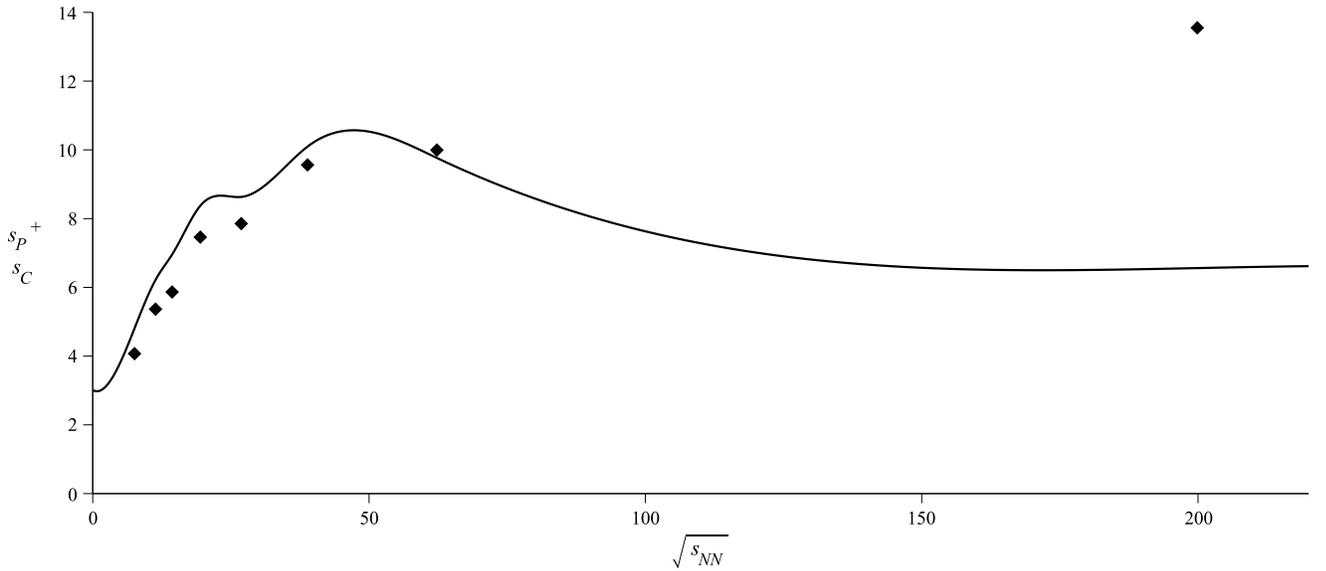}
\caption{Spline interpolation of the points in Figure 3, superimposed on Figure 4. Units on the vertical axis are (fm)$^{-3}$.}
\end{figure}

At low impact energies, then, the data for central collisions still respect the bound, despite the fact that the bound is reduced by vorticity; in other words, the vorticity has little effect at low impact energies. This is simply because in those cases (see Figure 2) the dimensionless quantity $\mathcal{A}^2/L^2$ is small. For example, for collisions at this centrality at 27 GeV impact energy, $\mathcal{A}^2/L^2 \approx 0.027$. To put it yet another way: for low impact energies (meaning $\sqrt{s_{\m{NN}}}$ at or below 39 GeV), the dual description by an AdS$_5$-Reissner-Nordstr\"om is adequate; we do not need to use an AdS$_5$-Kerr (or AdS$_5$-Kerr-Newman) metric in the bulk.

At the other extreme, impact energy 200 GeV, the upper bound drops dramatically, in fact to about half of the value of the entropy density given in \cite{kn:sahoo} for central collisions at this impact energy (as mentioned above, around 13.5 (fm)$^{-3}$). Thus clearly the model predicts that \emph{very large vorticities have the effect of sharply suppressing the entropy density}.

In fact, this is not unreasonable physically. It is thought \cite{kn:shub} that strong magnetic fields can dramatically constrict the plasma phase space, thus causing the entropy density of the QGP to fall. In view of the well-known close analogy between the effects of rotation and magnetism, it would not be surprising if the extremely large vorticities exhibited by plasmas produced in high-energy peripheral collisions can have similar consequences; and, in fact, the analogy may give a way of quantifying this effect. (In the language of complexity theory: one would use the analogy to compute the effect of vorticity on constricting the number of degrees of freedom available to ``execute gates''.) It will be interesting to see whether further investigations are able to confirm this prediction.

Note that our neglect of the baryonic chemical potential is justified at high impact energies: we compute $\mu_B/T \approx 0.14$ \cite{kn:phobos,kn:sahoo} in this case. That is, an AdS$_5$-Reissner-Nordstr\"om bulk is appropriate at low impact energies, but we can justify using an (uncharged) AdS$_5$-Kerr bulk at high impact energies.

This allows us to avoid having to use an AdS$_5$-Kerr-Newman bulk geometry, which, as mentioned earlier, is not actually known apart from various approximate metrics valid in certain restricted regimes. The one dubious case is $\sqrt{s_{\m{NN}}} = 62.4$ GeV, for which neither $\mu_B/T \approx 0.42$ nor $\mathcal{A}^2/L^2 \approx 0.09$ is completely negligible: notice that this is the impact energy at which vorticity pushes the upper bound down to approximate equality with the entropy density in the central case. We argued earlier that on theoretical grounds we do not expect the presence of a non-zero charge (or baryonic chemical potential) to affect the upper bound in any way in the rotating case, just as we know that it does not do so in the non-rotating case. If this expectation proves to be wrong, then collisions at $\sqrt{s_{\m{NN}}} = 62.4$ GeV will be extremely interesting and should be re-examined more closely.

\addtocounter{section}{1}
\section* {\large{\textsf{6. Conclusion: Complexity and the QGP}}}
The rate at which complexity grows in a field theory is thought to be related to the rate of growth of its entropy, and in fact it has been argued \cite{kn:second} that there is an explicit parallel between the second law of complexity and the second law of thermodynamics. In view of the existence of a (limited, approximate) holographic description of the QGP, it is therefore natural to ask whether the thermodynamics of the QGP might shed some light on the behaviour of complexity in general. In this work, we have proposed a concrete realisation of this idea: phenomenological models of the QGP support the idea that the product of its specific entropy density with its temperature, $sT/\varepsilon$, is bounded, at least for plasma in the strong-coupling regime produced in central collisions of heavy ions. This is in harmony with the fact that, for five-dimensional asymptotically AdS$_5$ black holes, the dual quantity, $ST/\mathcal{M}$, is also bounded. This latter bound depends on the angular momentum of the black hole, leading us to suggest that it will be found that the entropy density of the QGP is suppressed by extremely large vorticities.

That the quantity $ST/\mathcal{M}$ might be of interest was suggested by its conjectured relation with the rate of growth of the specific complexity. Bringing all of these threads together, one is led to ask: can we better understand the physical significance \cite{kn:suss5} of the growth of complexity, and, specifically, of the growth of the complexity of the QGP? The answers to such questions could give new insights into the problems of understanding strongly coupled plasmas in Nature.

\addtocounter{section}{1}
\section*{\large{\textsf{Acknowledgement}}}
The author is grateful to Dr. Soon Wanmei for useful discussions.

\end{document}